\documentclass[aps,prb,twocolumn,letterpaper,superscriptaddress,showpacs]{revtex4-1}
\usepackage{amsmath, amsthm}
\usepackage{keyval}
\usepackage{graphicx,latexsym}
\usepackage{dcolumn}
\usepackage{bm}
\usepackage{color}
\usepackage{url}
\usepackage{hyperref}
\usepackage{CJK}

\begin{document}
\title{Non-Fermi liquid scattering against an emergent Bose liquid: manifestations in the kink and other exotic quasiparticle behaviors in the normal-state cuprate superconductors}

\begin{CJK*}{UTF8}{}

\author{Shengtao Jiang (\CJKfamily{gbsn}蒋晟韬)}
\affiliation{Tsung-Dao Lee Institute \& School of Physics and Astronomy, Shanghai Jiao Tong University, Shanghai 200240, China}

\author{Long Zou (\CJKfamily{gbsn}邹龙)}
\affiliation{Tsung-Dao Lee Institute \& School of Physics and Astronomy, Shanghai Jiao Tong University, Shanghai 200240, China}
\author{Wei Ku (\CJKfamily{bsmi}顧威)}
\altaffiliation{corresponding author: weiku@mailaps.org}
\affiliation{Tsung-Dao Lee Institute \& School of Physics and Astronomy, Shanghai Jiao Tong University, Shanghai 200240, China}
\affiliation{Key Laboratory of Artificial Structures and Quantum Control (Ministry of Education), Shanghai 200240, China}

\date{\today}

\begin{abstract}
The normal state of cuprate superconductors exhibits many exotic behaviors qualitatively different from the Fermi liquid, the foundation of condensed matter physics.
Here we demonstrate that non-Fermi liquid behaviors emerge naturally from scattering against an emergent Bose liquid.
Particularly, we find a \textit{finite} zero-energy scattering rate at the low-temperature limit that grows linearly with respect to temperature, against clean fermions' generic nondissipative characteristics.
Surprisingly, three other seemingly unrelated experimental observations are also produced, including the well-studied ``kink'' in the quasiparticle dispersion, as well as the puzzling correspondences between the normal and superconducting state.
Our findings provide a general route for fermionic systems to generate non-Fermi liquid behavior and suggest that by room temperature large number of the doped holes in the cuprates have already formed an emergent Bose liquid of tightly bound pairs, whose low-temperature condensation gives unconventional superconductivity.

\end{abstract}

\maketitle
\end{CJK*}

\section{Introduction}

The Landau Fermi liquid (FL) theory is one of the corner-stones of condensed matter physics that explains most of the basic properties of materials~\cite{paper47}.
Another known generic liquid is the Luttinger liquid~\cite{paper48}, which seems to be another fixed point of interacting fermionic systems~\cite{paper49}.
However, in many strongly correlated materials, various non-Fermi liquid (NFL) behaviors have been observed experimentally that are qualitatively distinct from these known generic liquid behaviors.
The most well known example is the normal state of cuprate superconductors~\cite{paper3,paper2}.
The observed resistivity shows a linear temperature dependence in a wide doping and temperature range~\cite{paper7,paper8}.
This so-called bad-metal behavior is in great contrast to the generic quadratic dependence in the FL and has led to the suggestion of a ``hidden Fermi liquid''~\cite{paper53,dmft1}.
Furthermore, in the hole underdoped ``pseudogap'' regime, the Fermi surface becomes an open ``arc,'' beyond which the spectral function demonstrates an incomplete gap-like feature near momentum $k=(\pi,0)$~\cite{paper10,paper11}, without a well-defined quasiparticle peak~\cite{paper50,paper51,paper52}.

Even in the Fermi arc, where a quasiparticle-like peak can be observed by angule-resolved photoemission spectroscopy (ARPES), the peak is accompanied by a ``background'' spanning a large energy range taking a significant amount (at least half) of weight from the peak~\cite{paper3}.
Furthermore, in optimally doped samples, the scattering rate of the quasiparticles near the chemical potential is believed to have a $(\sqrt{T^2+\omega^2})$ like temperature $T$/energy $\omega$-dependence~\cite{paper31}.
This is very exotic, as it implies a nonanalyticity at the zero-temperature, zero-energy limit of the electronic self-energy, qualitatively different from the analytical $T^2+\omega^2$ dependence of the FL~\cite{paper33} and the $T+\omega^2$ dependence of the hidden Fermi liquid~\cite{paper53,hdl08}.
In fact, the smooth FL behavior has a profound origin related to the diminishing phase space of clean fermionic systems (not just FL) at low energy that reduces the scattering rate to zero in this limit.
(In other words, given the Pauli principle, clean fermionic systems are not supposed to have dissipation near the ground state.)
The observation implies unusual non-analytic behavior that perhaps further promotes the notion of a quantum critical point~\cite{paper31,paper55,paper56}, whose associated quantum fluctuation can in principle lead to unconventional superconductivity~\cite{paper57,paper58,paper59}.

This exotic scattering rate has been one of the most essential puzzles of condensed matter physics, in association with the above bad-metal behavior.
However, its microscopic origin remains elusive.
A phenomenological interpretation is the marginal Fermi liquid (MFL) which hypothesizes  charge and spin polarizabilities~\cite{paper24} from unknown physical origins.
More recently, the same nonanalytic behavior was shown to appear via holographic gauge/gravity duality~\cite{paper34,paper35}.
A realistic physical picture of this exciting new line of consideration still requires further development.

Even more unexpectedly, a recent experiment~\cite{paper23} found very similar structures in the high-temperature normal-state self-energy (which gives the scattering rate) and the anomalous self-energy in the low-temperature superconducting state (which gives the superconducting gap).
This is in excellent agreement with the earlier observation~\cite{paper36} that the normal-state quasiparticle scattering rate on the Fermi surface correlates directly with the low-temperature superconducting gap in multiple materials near optimal doping.
Together, these observations indicate that whatever constitutes the microscopic mechanism of superconductivity at low temperature, has already been encoded in the scattering of the normal state, a feature of the large energy scale of the essential correlations absent in all weak-coupling pictures.

In addition to these unusual behaviors that connect profoundly to the most basic concepts of condensed matter physics, and the recent studies on the charge-density wave~\cite{paper12,paper13,paper14,paper15}, quasiparticles in the cuprates present another universal and distinct ``kink'' in their dispersion~\cite{paper16,paper17,paper18,paper19,paper20,paper21,paper22}.
Coupling a MFL to the magnetic resonance in the superconducting state ~\cite{paper18,paper20,paper25} was proposed to be the origin of the kink, but the lack of magnetic resonance above the superconducting transition temperature, $T_c$ appears to contradict the observation of the kink above $T_c$~\cite{paper16,paper21}.
Coupling to the phonon~\cite{paper16,paper19} provides another possible origin, but its strength is questioned by a later calculation~\cite{paper26}.
A similar structure can also be produced by replacing the phonon by the spin fluctuation~\cite{paper27,paper28}, but no consensus has been reached for such a mechanism.
Notice however that none of these proposals properly includes the essential NFL scattering mentioned above that is obviously controlling the low-energy physics.

The combination of these four characteristics in the one-particle spectral function indicates unambiguously that the cuprates are in a many-body state \textit{completely} distinct from the usual Fermi liquid.
Then, other than a vague ``strongly correlated electronic system,'' what exactly are the cuprates?
The best-known attempt to answer this essential question is probably Anderson's ``hidden Fermi liquid''~\cite{paper53,hdl08}, which, however, does not naturally incorporate the above-mentioned strong correspondence between the superconducting gap and the normal-state scattering rate.

In this paper, we show that NFL scattering rate results naturally from scattering against an emergent Bose liquid of tightly bound pairs.
Near the optimal doping, we find a \textit{finite} scattering rate even at the zero-temperature and zero-energy limit that grows linearly with temperature, in contrast to the typical FL behavior.
In essence, the formation of bosonic pairs allows finite thermal fluctuation (and thus dissipation) in the low-temperature, low-energy limit, in the absence of condensation.
Note that such a NFL scattering rate is produced with an analytical self-energy and thus does not require a quantum critical point.
Most unexpectedly, the same scattering also produces a kink in the quasiparticle dispersion at the experimentally observed energy, revealing that the kink is essentially another manifestation of the underlying NFL scattering process.
Furthermore, our results give the observed direct correspondences between the normal and superconducting states in several cuprates, including their structures of the self-energies and scattering rate vs. superconducting gap.
Our study demonstrates a generic route for clean fermionic systems to break the fermionic zero-dissipation characteristics.
The simultaneous description of these seemingly unrelated experimental observations in the cuprates by a \textit{single} model suggests strongly that by room temperature a large number of the doped holes in the cuprates have formed an ``emergent Bose liquid'', whose condensation at low temperature gives the unconventional superconductivity.
\begin{figure}[h]
    \vspace*{-0.6cm}
    \includegraphics[width=1.1\columnwidth,clip=true]{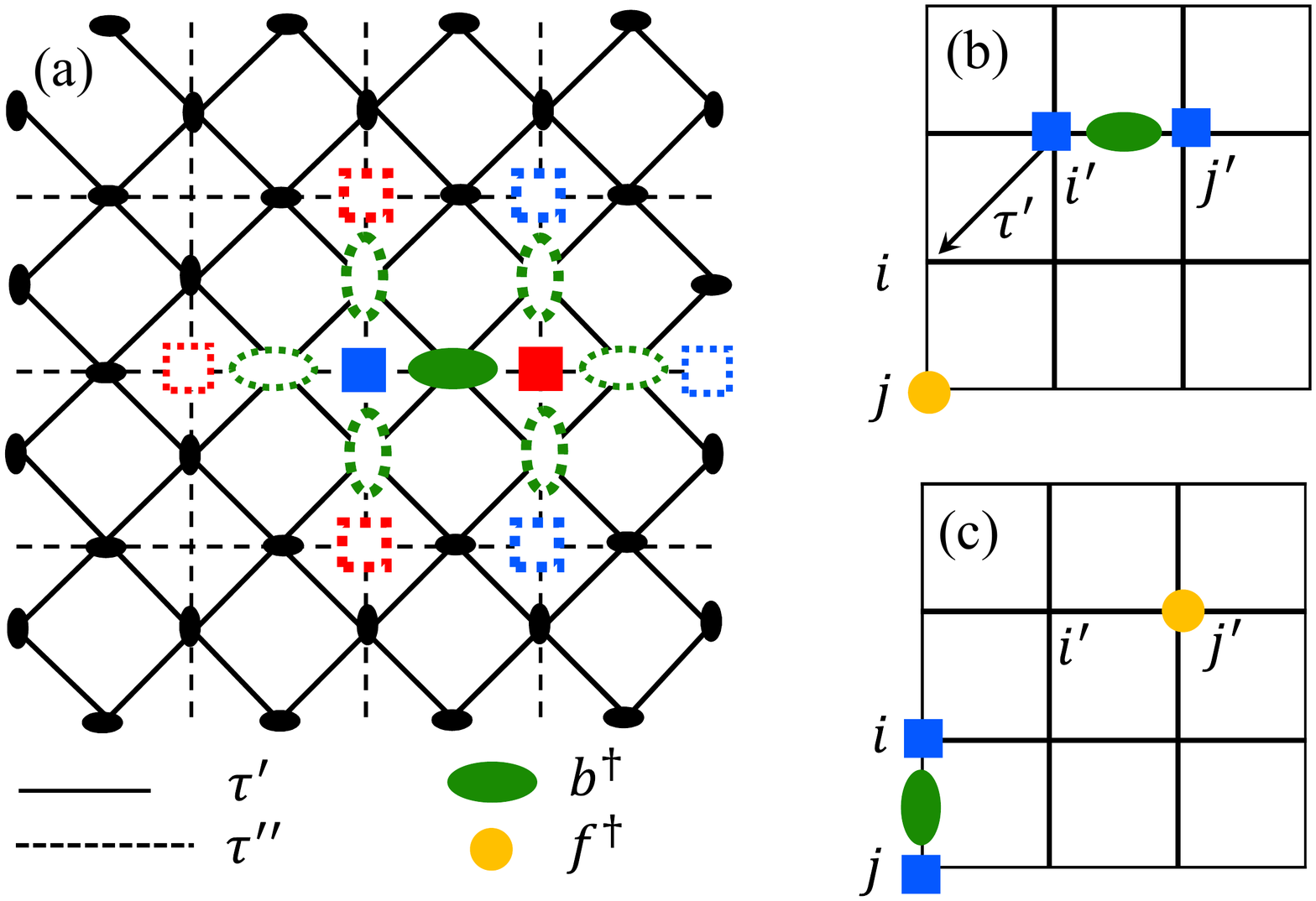}
	\vspace*{-1.2cm}
	\caption{\label{fig:fig0}
(a) Illustration of pivoting motion of EBL in Eq.~(\ref{eq6}).
The green solid ellipse denotes a bosonic tightly bound pair of holes located at the blue and red solid squares.
Through the second- and third- nearest-neighbor hoppings of holes (open squares), $\tau^\prime$ and $\tau^{\prime\prime}$, the boson can hop to the first- and second- nearest-neighbor bosonic sites (open ellipses).
The resulting bosonic lattice (black ellipses) forms a checkerboard lattice.
(b) and (c) Illustration of the scattering process $\tau_{ii^\prime}b^{\dagger}_{ij}f_{j}f^{\dagger}_{j^\prime}b_{i^\prime j^\prime}$, of a photohole (yellow circle) against a boson in Eq.~(\ref{eq1}).
	}
	\vspace*{-0.2cm}
\end{figure}
\section{ Model}
We assume a model system with very strong short-range correlations in spin, charge, and pairing channels, corresponding to energy scales much larger than the temperature and energy range of experimental interest.
In such a limiting case, these correlations would appear to be ``frozen'' or saturated in experimentally observed low-energy physics.
We further assume~\cite{paper37,paper43} concerning the charge and pairing degrees of freedom, the essential correlations manifest themselves to three constraints of the doped holes in the system: 1) no double occupancy of sites, 2) the formation of tightly bound nearest neighboring pairs of doped holes and 3) a fixed total number of bosons (since the pair-breaking fluctuation is assumed to be of higher energy and can be integrated out).
These assumptions lead to a simple model~\cite{paper37,paper43} of an emergent Bose liquid (EBL) in a checkerboard lattice (a two orbital Hamiltonian corresponding to the two types of neighboring bonds, vertical and horizontal) as shown in Fig.~\ref{fig:fig0}(a):
\begin{equation}
\label{eq6}
H^{b}=\sum_{ii^\prime, j\in \text{NN}(i)\cap \text{NN}(i^\prime)}\tau_{ii^\prime }b^{\dagger}_{ij}b_{i^\prime j},
\end{equation}
where $b_{i^\prime j}$ denotes the annihilation of a boson composed of fermions sitting at Cu site $i^\prime$ and its \textit{adjacent} site $j$.  $\tau_{ii^{'}}=\tau^\prime$ or $\tau^{\prime\prime}$ is the strength of a fully dressed kinetic process involving second- or third-nearest-neighbor sites, describing the pivoting motion of the two-legged boson.
The resulting non-interacting band structure and density of states of typical solutions of this two-orbital model are illustrated in Fig.~\ref{fig:fig3} below and correspond to the one-particle propagator of the boson, $D=1/(\omega-H^b)$~\cite{supp1}.

Justifications for applying this idealized model to the actual cuprates can be argued from general theoretical grounds~\cite{paper37,paper43} and are at least consistent with interpretations of many experimental observations~\cite{paper38,paper39,paper40,ong1,ong2,ong3,shi,bollinger2011}.
This model also takes into consideration the importance of phase fluctuation~\cite{doniach1990,paper41,paper42} for the superconductivity in the underdoped cuprates.
But of course, the ultimate justification for this model, particularly in contrast to other various proposals of ``preformed pairs''~\cite{preformed1,preformed2,preformed3,preformed4}, should come from verification of its physical properties against \textit{all} available experiments.
Previously, \textit{without using any free parameter}, this model successfully explained quantitatively the demise of superconductivity at 5.2\% doping~\cite{paper43} in excellent agreement with experiments, and produced a kinetics-driven second kind of superconducting gap with the correct experimental gap size~\cite{paper37}.
Below we will use this model to explain intuitively the novel physics behind all four main characteristics of the electronic spectral functions, giving further credibility to this model.

\begin{figure}[h]
	\vspace{-0.3cm}
	\includegraphics[width=1.0\columnwidth,clip=true]{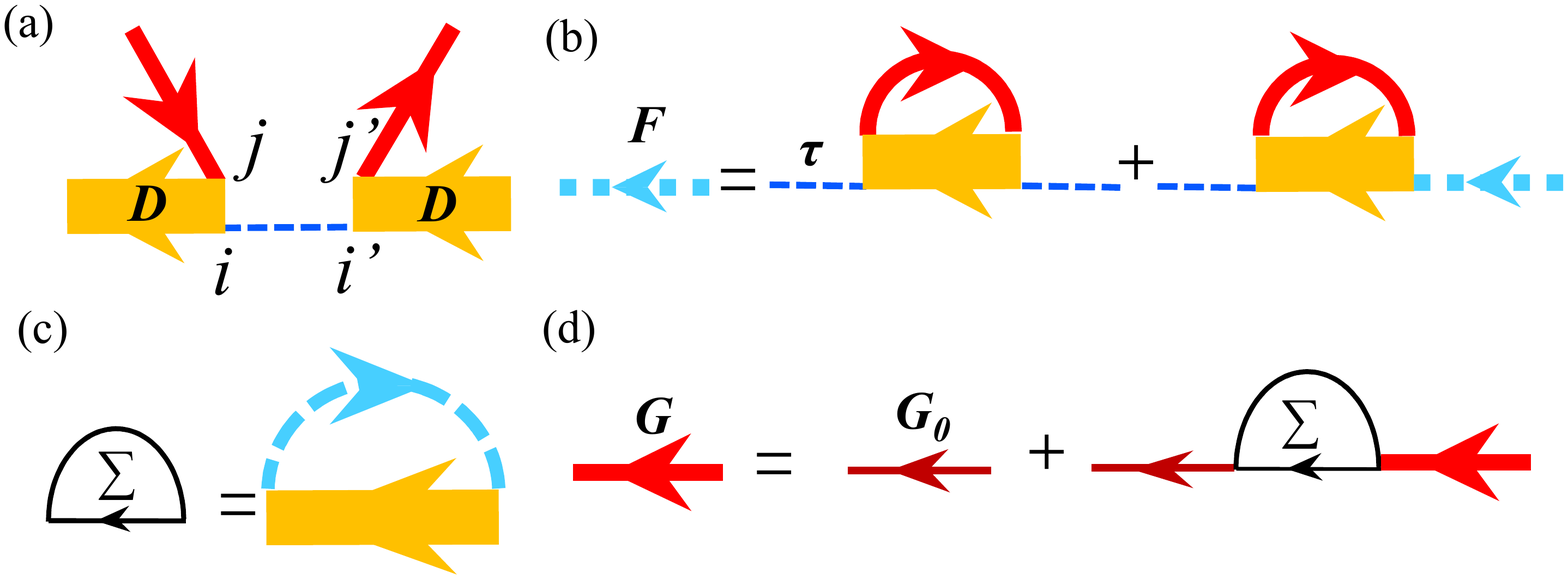}
	\vspace*{-3.0cm}
	\caption{\label{fig:fig1}
		Feynman diagram of (a) a kernel of quasiparticle scattering against two-orbital EBL, (b) Dyson's equation for dressed hopping $F$, (c) self-energy$\Sigma$, and (d) Dyson's equation for the dressed fermionic propagator $G$. The dotted blue line stands for bare hopping$\tau_{ii^{'}}$. The thick red line denotes the extracted renormalized propagator $G$.
	}
	\vspace*{0.2cm}
\end{figure}

We are most concerned about the effects on the electronic one-particle propagator $G$ of the injected photohole and the small number of residual unpaired holes (created by $f^\dagger$) via scattering against the bosonic pairs composed of holes \textit{indistinguishable} from them.
As an illustration, we consider the inelastic scattering process~\cite{paper43} that conserves the bosonic particle number, as shown in  Fig.~\ref{fig:fig0}(b) and (c):
\begin{equation}
\label{eq1}
{\sum_{ii^\prime}}{\sum_{\substack{j\in \text{NN}(i),\\ j^\prime\in \text{NN}(i^\prime)\setminus  j}}}\tau_{ii^\prime}b^{\dagger}_{ij}f_{j}f^{\dagger}_{j^\prime}b_{i^\prime j^\prime},
\end{equation}
Treating this process as a perturbation and making use of Wick's theorem, we derive the corresponding Feynman diagrams and their rules~\cite{supp2}.
We then perform the following partial sum of fermionic self-energy diagrams at finite temperature (see Fig.~\ref{fig:fig1})
\begin{equation}
\label{eq2}
\Sigma(1,1^\prime)=F(\overline{2^\prime},\overline{2})D(1,\overline{2};1^\prime,\overline{2^\prime}),
\end{equation}
[in the 1 $\rightarrow$ (space,time) notation, with variables with an overline denoting dummy ones to be summed over].
Here $D(1,2;1^\prime,2^\prime)$ denotes the propagation of the boson from $1^\prime$ and its adjacent $2^\prime$ to 1 and its adjacent 2. $F$ denotes the dressed hopping obtained from (in matrix notation)
\begin{equation}
\label{eq3}
F=\tau S\tau+\tau SF,
\end{equation}
which is dressed by
\begin{equation}
S(1,1')=G(\overline{2'},\overline{2})D(1,\overline{2};1',\overline{2'})
\end{equation}
via the dressed fermion one-particle propagator $G$, which itself is self-consistently obtained with the self-energy (in matrix notation)
\begin{equation}
\label{eq5}
G=G_{0}+G_{0}\Sigma G
\end{equation}
Note that in Eq.~(\ref{eq3}), the lowest-order term containing only the bare hopping is removed since its contribution to Eq.~(\ref{eq2}) leads to a nearly $k$ independent constant that can be absorbed by the chemical potential.

\begin{figure*}[t]
	\begin{center}
		\vspace{-2.0cm}
		\resizebox*{1.5\columnwidth}{!}{\includegraphics{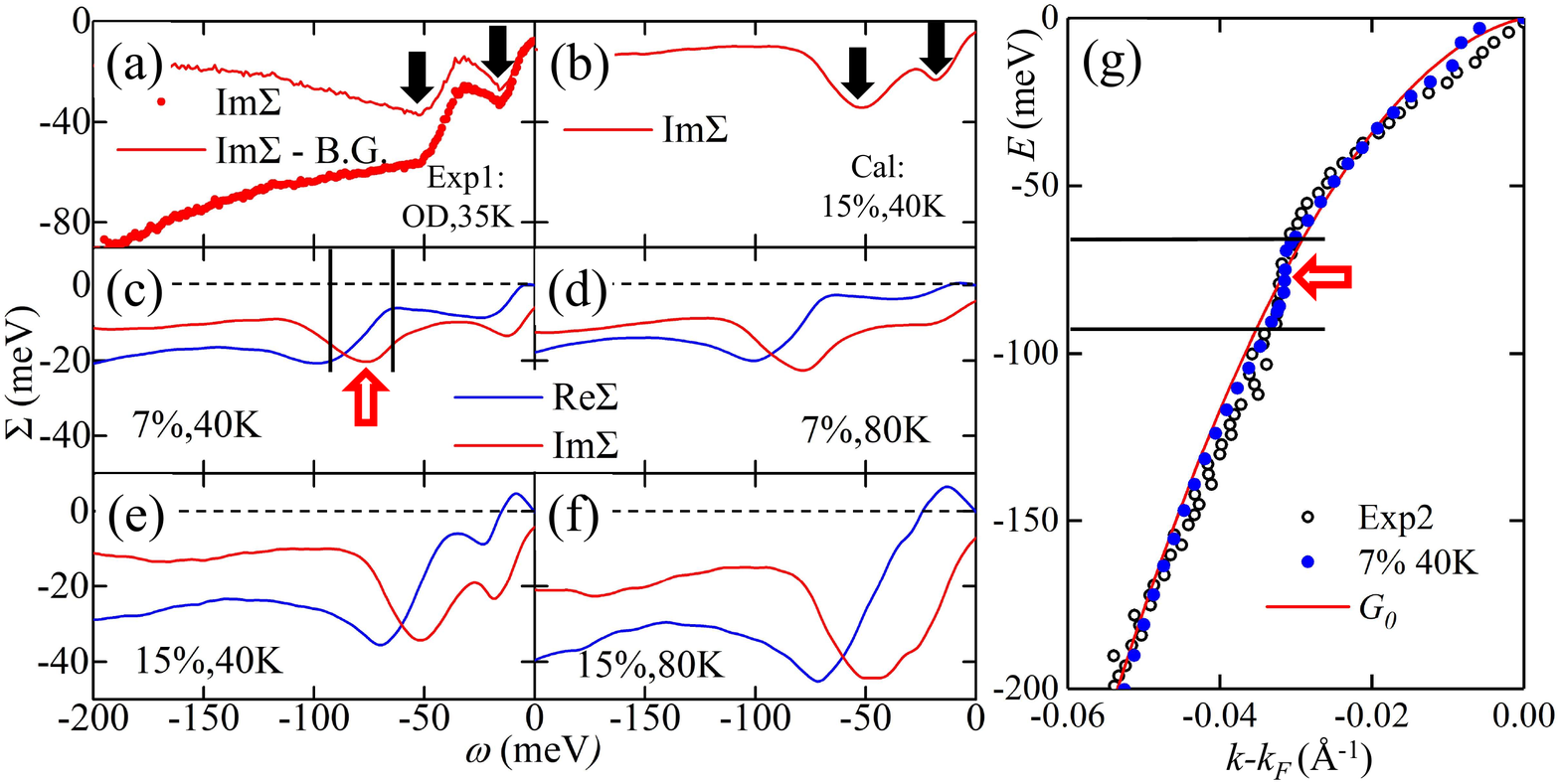}}
	\end{center}
	\vspace{-2.0cm}
	\caption{\label{fig:fig2}
		Real (blue) and imaginary (red) parts of the (a) experimental~\cite{paper23} and (b)-(f) calculated self-energy at different doping and temperatures.
		Solid arrows mark two distinct features in Im$\Sigma$.
		Open arrows and black lines indicate the direct correspondence of the peak feature and (g) the kink observed in the dispersion.
		The background removed in (a) is a quasilinear analytical $\sqrt{c^2+\omega^2}-c$ function with $c$ much smaller than the first feature around 20 meV, so that it does not introduce any visible artificial feature.
	}
	\vspace*{-0.4cm}
\end{figure*}

To best account for realistic cuprate materials, we use the same doping-dependent $\tau$ parameters in Refs.~~\cite{paper37,paper43} obtained from the dispersion $\epsilon$ near the chemical potential in the ARPES measurement of La$_{2-x}$Sr$_x$CuO$_4$.
We further use the same experimental dispersion to construct an approximate $G\sim \widetilde{G}(\vec{q},\omega)=W/[\omega-\epsilon (\vec{q})]$ with a reduced quasiparticle weight $W\sim 0.5$, roughly estimated from the experimental spectra~\cite{paper3} that shows a large weight loss in the incoherent features.
We then choose a featureless reference $G_0$ to ensure that the low-energy part of the resulting $G$ from Eq.(~\ref{eq5}) agrees well with $\widetilde{G}$ (experiment) to respect as much as possible the self-consistency of our formalism.
It is important to note that the third assumption above dictates that the chemical potential of the boson needs to be calculated for each temperature to guarantee the fixed particle number ($\sim x/2$) of bosons.

\section{Results}
Figure~\ref{fig:fig2} shows our calculated normal-state self-energy at two temperatures (40 and 80K).
Also shown in Fig.~\cite{fig:fig2}(a) is the measured Im$\Sigma$ with high resolution~\cite{paper23} for optimally doped Bi$_{2}$Sr$_{2}$CaCu$_{2}$O$_{8+\delta}$, which contains two distinct features at low energy (easier to see after removing a featureless background associated with other decay channels).
Amazingly, these features and, particularly, their energies are very nicely captured in our results in Fig.~\cite{fig:fig2}(b).
The agreement in their energies is not to be taken lightly, considering that it results from a framework that has \textit{no free parameter}: the essential parameters $\tau$ and $\widetilde{G}$ are obtained directly from ARPES experimental dispersion, and $D$ is obtained from $\tau$ directly.
(The weight reduction factor of 0.5 in $\widetilde{G}$ roughly estimated from the experiments mostly just fine-tunes the intensity of our results and does not alter their energies much.)

In fact, in our calculation, the stronger peak around 50 meV obtains its energy approximately from the binding energy of the Van Hove singularity at ($\pi$,0) given directly from the ARPES dispersion of La$_{2-x}$Sr$_x$CuO$_4$~\cite{paper37,supp3}.
(A similar energy of the Van Hove singularity was observed in optimally doped Bi$_{2}$Sr$_{2}$CaCu$_{2}$O$_{8+\delta}$ as well~\cite{paper3}.)
Consistently, Fig.~\ref{fig:fig2}(c)-~\ref{fig:fig2}(f) shows that at lower (7$\%$) doping the feature grows to 70meV, again following the well-known nonrigid band shift of the Van Hove singularity at ($\pi$,0)~\cite{paper44}.
Microscopically, this comes simply from the significant opening of the scattering phase space near the van Hove singularity.
This realization accounts naturally for the observed sudden change in momentum distribution curve of ARPES~\cite{paper16,paper17,paper18,paper19,paper21} as well.

Most unexpectedly, Figs.~\ref{fig:fig2}(c) and~\ref{fig:fig2}(g) show that when applied to a smooth featureless reference $G_0=W/[\omega-\epsilon_{0} (\vec{q})]$, this stronger feature produces a clear kink structure in the dispersion of the resulting $G$, a structure intensively studied by ARPES.
A careful examination of the resulting kink structure should make clear that both experimental dispersion and our calculated dispersion actually contain two kinks (marked by thin lines), between which the dispersion is steeper.
This behavior is also clearly observed in experimental data shown here (6.3\%, 20K)~\cite{paper50} and in other experiments~\cite{paper20,paper21,paper22}.
Such a flat-steep-flat, two-kink structure is qualitatively distinct from the flat-vertical, one-kink structure produced by coupling to phonon~\cite{paper26} and spin fluctuations~\cite{paper28}, as it requires a peak, not a dip, in Im$\Sigma$.
Since the kink energy is closely related to the Van Hove singularity derived peak in Im$\Sigma$, one should expect a systematic correspondence between these two measured quantities.
Indeed, in overdoped Bi$_{2}$Sr$_{2}$Cu$_{2}$O$_{8+\delta}$, the Van Hove singularity occurs at higher energy around 100meV~\cite{feng2001}, and correspondingly, a kink of the similar energy was reported by ARPES~\cite{paper20}.

\begin{figure*}[t]
	\begin{center}
		\vspace*{-2.4cm}
		\includegraphics[width=1.8\columnwidth,clip=true]{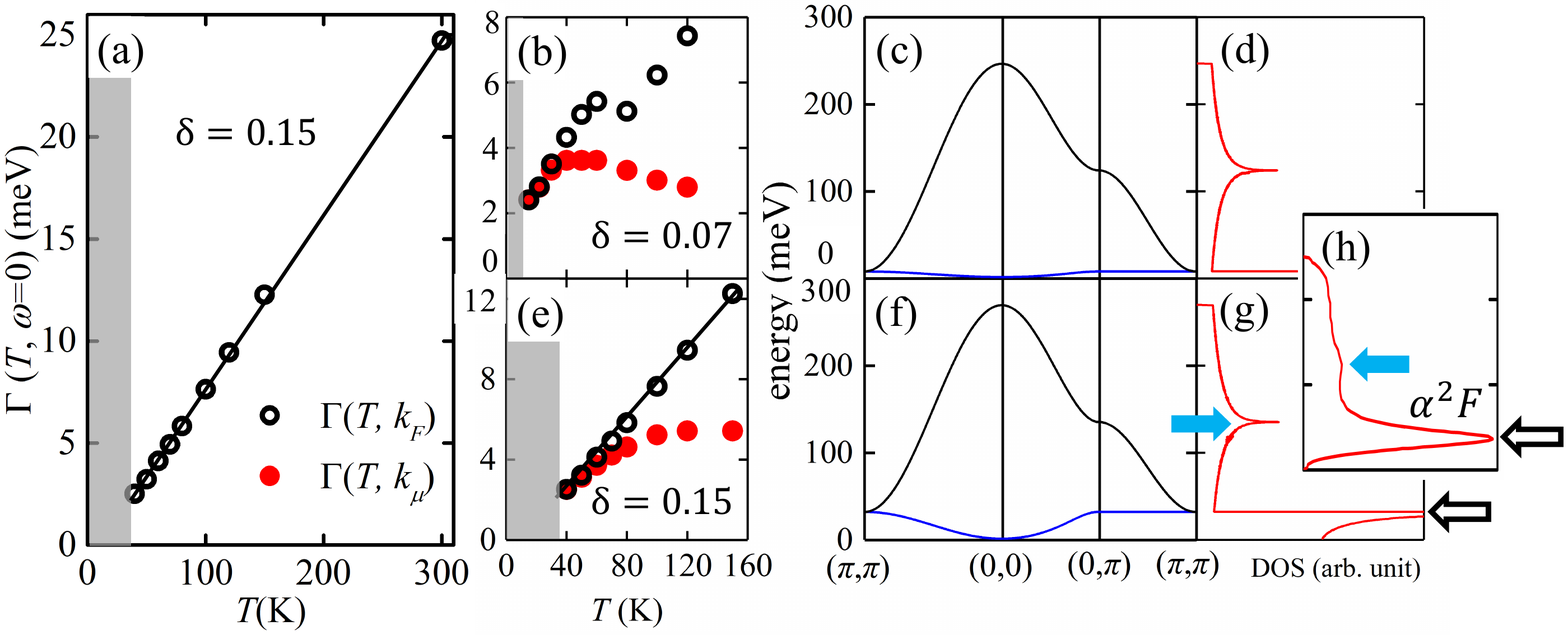}
	\end{center}
	\vspace*{-3.0cm}
	\caption{
		\label{fig:fig3}
		(a), (b), (e) Temperature-dependent scattering rate $\Gamma$ of the nodal quasiparticle at the Fermi wavevector $k_F$ (black open circles) and at the chemical potential $k_\mu$ (red solid circles) for $7\%$ (top panel) and $15\%$ (bottom panel).
Both show NFL behavior with a linear temperature dependence with a constant at the zero-temperature limit. The shaded region is below $T_C$.
(c) and (f) The corresponding band structure and (d) and (g) densitys of state of the bosonic pairs. (h) The experimentally extracted Eliashberg function $\alpha^{2}F(\omega)$~\cite{paper23} at optimal doping, showing a strong peak around 40meV, a weak peak around 140meV, and a large energy range around 250meV, all of which are well captured by the calculated bosonic density of states in (g).
	}
	\vspace{-0.3cm}	
\end{figure*}

We now show that near the optimal doping, this model produces an exotic NFL scattering rate at low temperature.
Figure~\ref{fig:fig3}(a) shows a linear temperature $T$ dependent scattering rate $\Gamma(k_F)$ measured from the full width at half maximum of the peak in our resulting spectral function at the fixed Fermi wave vector $k_F$.
Such a linear temperature dependence signifies an exotic scattering, qualitatively distinct from the standard $T^2$ dependence scattering of a FL.
This linear dependence has been observed experimentally near the optimal doping~\cite{paper29,paper31,kondo2015}, and is regarded as the phenomenological MFL~\cite{paper24}.

Even more exotically, Fig.~\ref{fig:fig3}(b) shows that the scattering rate approaches a \textit{finite} value at the low temperature, indicating that the low-energy carriers can dissipate even at low-temperature limit without disorder.
This is quite unexpected, since generally speaking, due to the Pauli principle, a typical \textit{clean} fermionic system should have a diminishing phase space of scattering at zero temperature and cannot dissipate at the chemical potential.
This is why even the phenomenological MFL picture assumes a zero Im$\Sigma$ at the chemical potential as temperature approaches zero and why the hidden Fermi liquid shows the same behavior.
This is also the reason why such a finite scattering rate is always ignored in experimental analysis~\cite{paper21,paper23} by regarding disorder as its origin.
Our results open an entirely new possibility that such finite scattering might be intrinsic to the clean fermionic system, and should be analyzed with care in future experiments. 

This issue has a significant physical consequence.
If, indeed, the scattering rate must be zero at the chemical potential, the observed linear $\omega$ dependence necessarily dictates a non-analytical function of $\omega$.
Such non-analytical behavior, is of course, quite special and might support the notion of a quantum critical point~\cite{paper31,paper55,paper56}, for example.
However, if the scattering rate is allowed to be finite at the chemical potential, as found here, a linear $\omega$ dependence comes simply from the lowest-order expansion of an analytical function, $\text{Im}\Sigma(\vec{k}_{F},\omega ,T)\approx a_{0}+a_{1}T+a_{2}\omega$.

So how can our model break the above generic phase-space limitation of dissipation of fermions?
The answer lies in the nontrivial EBL.
On the one hand, the indistinguishableness between the photohole and the holes that constitute the boson results in scattering processes like that in Eq.~(\ref{eq1}).
On the other hand, the peculiar nature of larger thermal fluctuation of uncondensed bosons would produce incoherent scattering even at the zero-temperature limit.
In essence, by forming an EBL, the fermionic system can escape from its fermionic constraints.
Note that this consideration is clearly very general and does not rely on the details of our specific model.
Such NFL behavior is possible only in the limitless richness of emergence in many-body systems.
\begin{figure*}[t]
	\begin{center}
		\vspace{-2.0cm}
		\resizebox*{1.7\columnwidth}{!}{\includegraphics{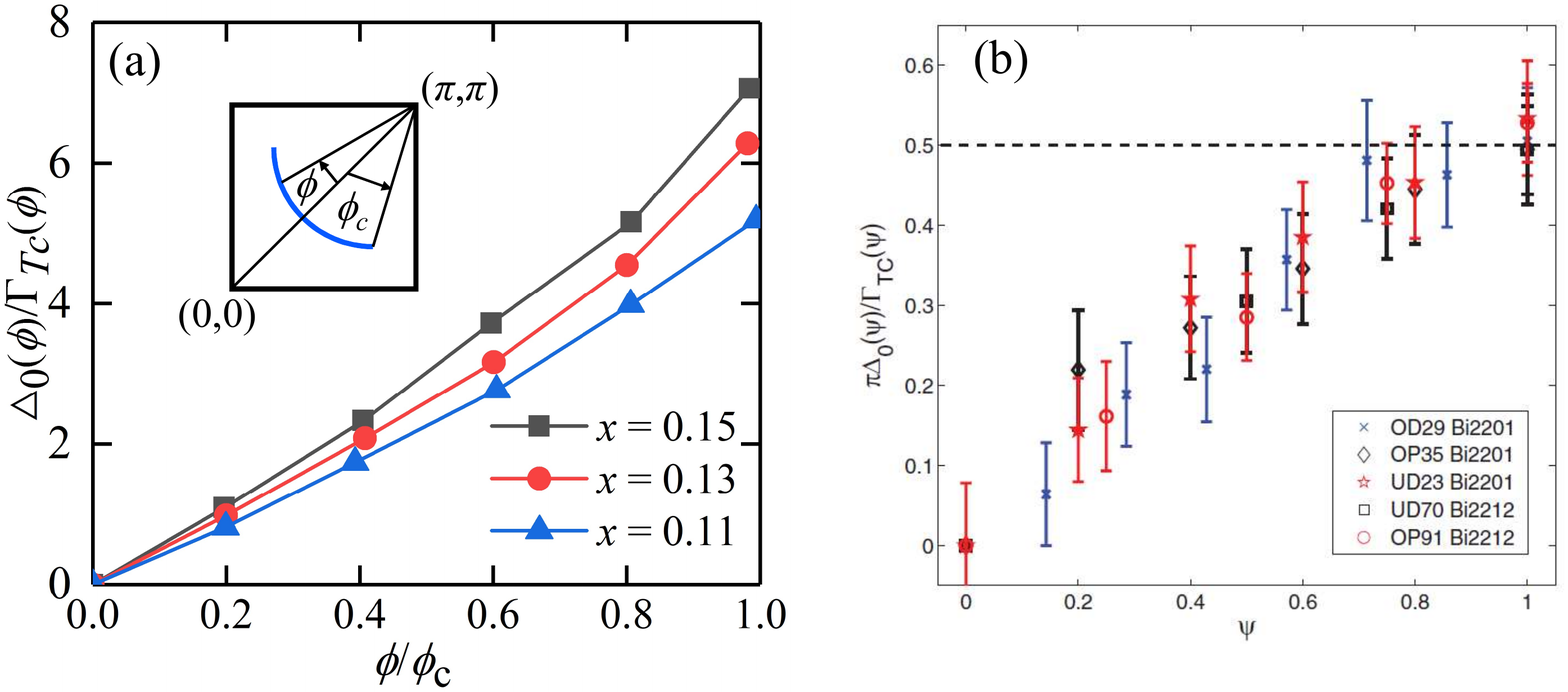}}
		\vspace{-2.7cm}
	\end{center}
	\caption{
		\label{fig:fig4}
		A scaling relationship between the kinetic-driven superconducting gap~\cite{paper37} at zero temperature and quasiparticle scattering rate at the normal state slightly above $T_{C}$ in La$_{2-x}$Sr$_x$CuO$_4$, as a function the of rescaled Fermi surface $\phi/\phi_{c}$. (a) The calculated trend resembles (b) the observed one in Bi series cuprates ~\cite{paper36}.
		The inset in (a) is an illustration of the Fermi surface angle $\phi$, with $\phi_{c}$ being the angle of the endpoint on the Fermi surface.}
	\vspace{-0.4cm}
\end{figure*}

The linear temperature dependence of the scattering rate can be visualized by rewriting Eq.~\ref{eq2} into approximately the form of the Eliashberg function:
\begin{equation}
\text{Im}\Sigma(\vec{k},0,T)=-\int \alpha^{2}F(\vec{k},u)[n_{b}(u,T)+n_{f}(u,T)]du,
\end{equation}
where the Eliashberg function $\alpha^{2}F$ is approximately proportional to the bosonic density of states (DOS).
The $n_b$-related first term yields a constant (since the number of tightly bound pairs is fixed), while the $n_f$-related second terms yields a linear temperature dependence~\cite{supp4}, as long as the DOS does not change too fast at the energy scale of $k_{B}T$.
This is why such linearity persists longer in optimally doped system, in which the band width of the lower-energy band is the biggest [see Fig.~\ref{fig:fig3}(g)].

Of course, this approximate analysis is limited to low temperature, where the chemical potentials for the boson and the photohole do not shift strongly with temperature.
Otherwise, Fig.~\ref{fig:fig3}(e) shows that at high temperature, the spectral function that peaks at the chemical potential will occur at a different wave vector $k_\mu$ and experience a different scattering rate.
Particularly, the shift of the bosonic chemical potential will cause the scattering channels to deviate from linear increase.
Interestingly, a similar reduction of the quasiparticle scattering rate at high temperature was produced by a dynamical mean-field calculation recently~\cite{dmft1}, even though the physics in play is quite different.

Finally, our picture also offers a natural explanation of the puzzling correspondence between the normal-state scattering rate and the superconducting gap~\cite{paper36}.
For example, ARPES measurements reported an unexpected correlation between the normal-state scattering rate at the transition temperature $\Gamma_{T_c}$ and the low-temperature superconducting gap $\Delta_0$ shown in Fig.~\ref{fig:fig4}(b):
\begin{equation}
\label{eq8}
\Delta_{0}(\phi)\propto \frac{\phi}{\phi_{c}} \Gamma_{Tc}(\phi),
\end{equation}
where $\phi$ denotes the $k$-space angle from the nodal point ($\pi$,$\pi$)/2, and $\phi_c$ denotes the same for the end of the Fermi arc [see the inset in Fig.~\ref{fig:fig4}(a)].
In the traditional weak-coupling theory of superconductivity, the superconducting gap is controlled by the strength of pairing, which does not leave much of a signature in the normal state when amplitude fluctuation overwhelms the system.
So, this correspondence is quite unimaginable in the weak coupling regime.

In our picture, on the other hand, this is quite straightforward.
As reported in a previous study~\cite{paper37}, at low temperature, a second kind of superconducting gap appears in the one-particle spectral function through coherent kinetic scattering against the condensed EBL.
This ``superconducting gap'' is a simple analytical function of the condensation density ($\sim x/2$, half of the doping level at zero temperature) and fully renormalized hopping $\tau_{ii^\prime}$.
With a $d$-wave condensate~~\cite{paper43}, the momentum dependence becomes simply linear near the nodal point: $\Delta_0 \propto \sqrt{x} \phi$.
On the other hand, the normal-state scattering rate results from inelastic scattering against the \textit{same} set of bosons, except they are not yet condensed, $\Gamma \propto x$ with very weak $k$ dependence due to the heavy convolution in Eq.~(\ref{eq5}).
Given that $\phi_{c}$ is approximately proportional to the Fermi arc length, which scales as the square root of the hole pocket size that is proportional to the doping level $x/2$, $\phi_c \propto \sqrt{x}$, the observed trend is easily understood.
Indeed, our results shown in Fig.~\ref{fig:fig4}(a) reproduce very nicely the observed trend of Eq.~(\ref{eq8}) in Fig.~\ref{fig:fig4}(b).

In essence, in this picture, all the short-range correlations are so strong that they become frozen at low temperature, including at the normal state slightly above $T_c$.
In other words, all the relevant information concerning the lower-temperature condensed state is already available in the normal state.
Therefore, this kind of direct correspondence between many properties of the normal state and the superconducting state is natural.

This consideration immediately applies to yet another observed correspondence.
The Eliashberg function $\alpha^2F$ of the normal and pairing self-energies extracted from high resolution laser ARPES data was found to have the same characteristics~\cite{paper23} [see Fig.~\ref{fig:fig3}(h)]: a strong peak around 40meV, a weak peak around 140meV, and a broad feature extending to 250meV.
(Calculation of conductivity data~\cite{paper46} also suggests such a large cutoff.)
It is obviously very hard to imagine a phonon extending to such a high energy, or spin fluctuation demonstrating such a rich structure.
However, compared to the DOS of our boson [see Fig.~\ref{fig:fig3}(h)], one immediately recognizes the resemblances in all three characteristics.
Again, both states are scattering against the \textit{same} set of boson, condensed or not.

\section{Summary}
In summary, we showed that the non-Fermi liquid scattering rate results naturally from scattering against an emergent Bose liquid of tightly bound pairs, designed to model the hole-doped cuprates.
At the chemical potential, even clean fermionic systems develop a \textit{finite} scattering rate at the zero-temperature limit that grows linearly with temperature, in contrast to the usual non-dissipative fermionic characteristics.
Such exotic behavior does not involve a non-analytic self-energy and does not require proximity to a quantum critical point.
Unexpectedly, the same non-Fermi liquid scattering process also generates a kink structure in the resulting one-particle propagator at the experimentally observed energy, revealing that the kink is another manifestation of the non-Fermi liquid scattering.
Our results further produced the observed direct correspondence between the normal-state scattering rate and the superconducting gap, as well as their underlying structures in the self-energy.
Our findings provide a generic route for fermionic systems to demonstrate non-Fermi liquid behavior. They also suggest that the cuprates are in this exotic regime in which a large number of doped holes develop bosonic features by forming an emergent Bose liquid of tightly bound pairs that condense into a superfluid at lower temperature.

We thank V. Dobrosavljevic, A. Hegg and S. Sen for useful discussions. Work was supported by National Natural Science Foundation of China  Grants No. 11674220 and No.  11447601, and Ministry of Science and Technology Grants No. 2016YFA0300500 and No. 2016YFA0300501.

\bibliography{cuprate}

\end{document}